\documentclass[a4paper,12pt]{article}      
\usepackage{graphicx}
\usepackage{amsmath}
\usepackage{amsfonts}

\begin{document}

\begin{flushright}			
       CHIBA-EP-134\\
       HEP-TH/0201086\\
       \tiny{Compiled \today}
\end{flushright}
\begin{center}

\renewcommand{\thefootnote}{\fnsymbol{footnote}}

{
\Large \sffamily \bfseries%
  {%
Integer- and Non-Integer-Shift of
the Chern-Simons Coupling under
a Local Higher Covariant Derivative Regulator
  }
}

\vspace{18pt}
{\large \sffamily \bfseries {Koh-ichi Nittoh}}
      \footnote{E-mail: \texttt{tea@cuphd.nd.chiba-u.ac.jp}}
\\
{\small\it
Graduate School of Science and Technology, Chiba University,\\
1-33 Yayoi-cho, Inage-ku, Chiba 263-8522, Japan}\\
\end{center}\vspace{18pt}
\begin{abstract}
The Chern-Simons coupling shift is calculated
within the framework of the hybrid regularization
based on a local higher covariant derivative regulator.
When the Yang-Mills term is present in the theory
the well-know integer-shift is obtained,
but is absent, the shift value is non-integer.
These results
show a possibility
that a non-integer-shift can be derived
using a local higher covariant derivative
and also suggest that the Yang-Mills term plays an important role
in the integer-shift of the Chern-Simons coupling.
\end{abstract}
\vspace{12pt}
{
}
\vspace{18pt}

\renewcommand{\thefootnote}{\fnsymbol{footnote}}

\label{sec:introduction}

After the Witten's paper
conjectured the Chern-Simons (CS) coupling shift
by a quantum effect in non-perturbative approach~\cite{Witten:1989hf},
many papers were published discussing the shift problem
in the view of a perturbative approach%
~\cite{Alvarez-Gaume:1990wk,Korchemsky:1990va,Martin:1990xv,Asorey:1990rn,
       Birmingham:1990ut,Giavarini:1993xb},
some papers gave the same result with Witten's but some were disagreement.
After a subsequent analysis with the geometrical regularization
by Asorey \textit{et al.}~\cite{Asorey:1994em},
it was clarified that
the shift is determined by the large momentum leading term of the regulator.

According to this paper, 
when we write the CS coupling shift as
$\theta\rightarrow\theta + \alpha$,
the shift value of $\alpha$ is classified
by the parity of the large momentum leading of the regulator
into three ``universality classes'':
(i) \textit{integer-shift}; $\alpha = c_v$ when the large-momentum leading of the regulator is parity-even,
(ii) \textit{no-shift}; $\alpha = 0$ when parity-odd%
\footnote{More precisely, $\alpha = 2sc_v$
          with $s$ a number of the pole of the function $\phi (p)$
          which is defined as the ratio between parity-even
          and -odd terms in the regularized action~\cite{Asorey:1994em}.
          In Ref.~\cite{Nittoh:1998ey},
          the function $\phi (p)$ has no pole
          so that only $\alpha = 0$ is obtained.
          }
and
(iii) \textit{non-integer-shift}; $\alpha = r c_v$ with a real number $r$
when a linear combination of parity-even and -odd terms.

These results were reanalyzed by a hybrid regularization with
higher covariant derivative (HCD) and dimensional regularization%
~\cite{Giavarini:1994zh},
it was shown that the results (i) and (ii) are realized
changing the parity of the regulator
which dominates in the large momentum limit.
They also argued from a dimensional analysis of the regulator
that the result (iii), however, does not realized in their method
because the large momentum leading term of the regulator
does not give a marginal state of parity-odd and -even.
This shows that,
even by our regularization method mentioned below,
the result (iii) cannot be reproduced 
as long as a local HCD term is used in the regularization scheme.

The dimensional regularization in the CS gauge theory is
problematic in the treatment of the antisymmetric symbol
$\epsilon^{\mu\rho\nu}$.
We use a Pauli-Villars (PV) regularization
instead of the dimensional one
in the hybrid regularization in this paper,
and show that unclassifiable result into (i) or (ii)
can be derived by a local HCD term,
which seems to belong (iii).


The regularization method of this paper is
a hybrid method of HCD and PV formulated in
Ref.~\cite{Nittoh:2000it} for the four--dimensional Yang-Mills (YM) theory.
This method is applicable for three--dimensions as follows.
%

Let us consider an SU($N$) gauge theory on $\mathbb{R}^3$
with an euclidean metric $g_{\mu\nu}$ which has a signature $(+++)$.
The SU($N$) gauge field is denoted by $A_\mu = A_\mu^a T^a$,
where $T^a$ is an anti-hermitian generator of
the Lie algebra and satisfies relations
$\left[T^a,T^b\right] = f^{abc}T^c$ and 
$\mathrm{Tr}\left(T^aT^b\right)=\frac{1}{2}\delta^{ab}$,
where $f^{abc}$ is a structure constant
completely anti-symmetric in its indices.

The generating functional regularized by this method is given by%
\begin{equation}
Z=
\int
\mathcal{D}A_\mu\,\mathcal{D}b\,
\mathcal{D}\overline c\,\mathcal{D}c\,
\exp \left[-S_\Lambda \right]
\prod_j \det{}^{-\frac{\alpha_j}{2}}\mathbf{A}_j
\prod_i \det{}^{\gamma_i}\mathbf{C}_i
\label{eq:generating functional}
\end{equation}
where $S_\Lambda$ is the action regularized by HCD,
and the determinants denoted by $\det{}^{-\frac{\alpha_j}{2}}\mathbf{A}_j$
and $\det{}^{\gamma_i}\mathbf{C}_i$ are
the PV determinants for gauge and ghost, respectively.

Introducing an infinite number of PV fields along this method,
we can construct a parity-invariant PV regulator.
Since such a regulator does not give any parity-odd contributions
to the quantum correction,
the source of the shift is uniquely identified to the HCD regulator.

When this method is used for the CS gauge theory with a parity-odd HCD term,
we have already shown in Ref.~\cite{Nittoh:1998ey},
the CS shift does not occur
because all the terms in the regularized action
have the parity-odd character.
This result is classified into (ii).
With a parity-even HCD, on the other hand,
some parity-even terms appear in the regularized action
and an odd-number product of parity-odd terms
can be constructed when we write one-loop diagrams.
Such term is expected to give a source of the shift.

\label{sec:YMCS}

First we consider the Yang-Mills-Chern-Simons (YMCS) theory
where a YM term is introduced as a part of HCD regulators.
The CS theory is recovered in an infinite limit of the parameter
$\mu\rightarrow\infty$.
%
The regularized action is given by
\begin{equation}
S_\Lambda = S_\mathrm{CS} + S_\mathrm{YM}
 + S_\mathrm{HCD} + S_\mathrm{GF}^H.
\label{eq:regularized YMCS action}
\end{equation}
$S_\mathrm{CS}$, $S_\mathrm{YM}$, $S_\mathrm{HCD}$ and $S_\mathrm{GF}^H$
are the actions of the CS, YM, HCD and gauge-fixing,
and their explicit forms are given by the following:
\begin{equation}
S_\mathrm{CS} =
-\mathrm{i}
\int \mathrm{d}^3 x 
\epsilon^{\mu\rho\nu}
\left(
\frac{1}{2}A_\mu^a \partial_\rho A_\nu^a
+\frac{1}{3!}gf^{abc}A_\mu^a A_\rho^b A_\nu^c
\right),
\label{eq:CS action}
\end{equation}
\begin{equation}
S_\mathrm{YM} =
\frac{1}{ 4\mu}
\int \mathrm{d}^3 x 
\left( F_{\mu\nu} \right)^a
\left( F^{\mu\nu} \right)^a,
\label{eq:YM action}
\end{equation}
\begin{equation}
S_\mathrm{HCD} =
\frac{1}{ 4\mu \Lambda^2}
\int \mathrm{d}^3 x 
\left( D_\lambda F_{\mu\nu} \right)^a
\left( D^\lambda F^{\mu\nu} \right)^a,
\label{eq:HCD action}
\end{equation}
\begin{equation}
S_\mathrm{GF}^H =
\int \mathrm{d}^3 x
\left[
 \frac{\xi_0 }{ 2} b^a b^a
 -
 b^a H(\partial/\Lambda) (\partial^\mu A_\mu)^a
 +
 \overline c^a H(\partial/\Lambda) 
 (\partial_\mu D^\mu c)^a
\right].
\label{eq:GF action}
\end{equation}
Here we take the gauge-fixing action $S_\mathrm{GF}$
as the same as one of the four--dimensional YM theory.
The fields $c$, $\overline c$ and $b$
are the ghost, anti-ghost and auxiliary field
and have the mass dimension $\frac{1}{2}$, $\frac{1}{2}$ and $1$ 
respectively.
The gauge-fixing parameter $\xi_0$ has the mass dimension one.
$H(\partial / \Lambda)$ is employed
to improve the convergence of the gauge-variant part of the propagator
which is most generally taken to be 
$H^2 = 1 + \frac{p^2}{\Lambda^2}$ in momentum space.
%

%
The explicit form of $S_\mathrm{HCD}$ is not exactly the same as 
the four--dimensional YM theory.
Suppose we take the HCD action in general form as
$\frac{1}{\mu \Lambda^{2n}}\int\mathrm{d}^3x(D^nF)^2$,
the superficial degree of divergence $\omega$
in three--dimensions is written by
$\omega = 3- (2n+1)(L-1)-E_A-(n/2+1)E_c$,
where $L$ is the number of loops, 
$E_A$ and $E_c$ are the numbers of external lines of
gauge and ghost fields respectively.
For $L\ge 2$, $n=1$ always gives a negative $\omega$
with some external legs.
%
Though the HCD action can be chosen to have the parity-odd character
as we introduced in the CS gauge theory~\cite{Nittoh:1998ey},
here we take the parity-even HCD action.
Such a HCD action is taken to be similar to
one of the four--dimensional YM theory
we can calculate the one-loop diagrams
with a minor modification of the result given by Ref.~\cite{Nittoh:2000it}.

All the diagrams higher than two are regularized by the above HCD action
but the two- and three-point functions at one-loop level are still divergent.
We use the PV method to regulate the remaining diagrams.

The PV determinants are denoted by
$\det{}^{-\frac{\alpha_{j}}{2}}\mathbf{A}_{j}$ 
and $\det^{\gamma_{i}}\mathbf{C}_{i}$
in \eqref{eq:generating functional}
where $i$ and $j$ run from $-\infty$ to $+\infty$ except zero.
Explicit forms of them are almost the same as the PV determinants
introduced in Ref.~\cite{Nittoh:2000it},
but slightly different in $S_{M_j}$ of
$\det{}^{-\frac{\alpha_{j}}{2}}\mathbf{A}_{j}
=\int \mathcal{D}A_j\, \mathcal{D}b_j\,
\mathrm{e}^{-S_{M_{j}} -S_{b_{j}} }$;
\begin{equation}
S_{M_j}=
\frac{1}{2}\int \mathrm{d}^3 x\, \mathrm{d}^3 y
A_j{}_\mu(x)
\left[
 \frac{\delta^2 S_{\Lambda}^{\mathrm{sgn}(j)}}
 {\delta A_\mu(x) \delta A_\nu(y)}
 - M_{j} \delta^{\mu\nu}\delta(x-y)
\right]
A_j{}_\nu(y),
\end{equation}
where $S_{\Lambda}^{\mathrm{sgn}(j)} \equiv
\mathrm{sgn}(j) S_\mathrm{CS} + S_\mathrm{YM}
 + S_\mathrm{HCD} + S_\mathrm{GF}^H$,
$\mathrm{sgn}(j)$ gives the signature of $j$
and $M_j\equiv M|j|$.
This determinant gives a parity-invariant PV pair
when $A_{+j}{}_\mu$ and $A_{-j}{}_\mu$  are exchanged
under the parity transformation%
~\cite{Nittoh:1998ey}.


Now we derive the value of the CS coupling shift
calculating the one-loop diagrams.
As we see in the above formulation,
almost terms appear in this theory are the same as
in the four--dimensional YM theory~\cite{Nittoh:2000it}
except appearance of the CS term.
So our task will be easily completed
by considering the distinctive point between
YMCS theory and pure YM theory.

The most different point from the four--dimensional YM theory is
the existence of the CS term
which only contributes to the Feynman rules of the gauge field:
the gauge propagator $\left<AA\right>$
and the three-point vertex $\left<AAA\right>$
(and also its PV field $\left<A_jA_j\right>$
and $\left<AA_jA_j\right>$) are modified.
These propagators and vertices obtain a parity-odd term
caused by the CS term,
so new contributions appear in one-loop corrections.
These contributions contain both the parity-odd and -even ones:
an odd-number product of parity-odd terms gives a parity-odd correction
and an even-number product a parity-even one.
%

According to the above discussion,
we only have to recalculate the diagrams containing a contribution
from the CS action.
They are diagrams of (a), (b), (c), (d), (i) and (j) in Fig.~1
of Ref~\cite{Nittoh:2000it}.
We concentrate on the corrections originated from the CS action
in the following discussions.
%
%
The diagrams of (a), (b), (i) and (j) give
both the parity-odd and -even corrections
because each propagator and vertex has a parity-odd term
so both the odd- and even-number products of parity-odd terms
can be derived.
%
On the other hand,
diagrams (c) and (d) give only parity-odd corrections
since the CS action does not have an effect on the four-point vertices
$\left< AAAA \right>$ and $\left< AAA_jA_j \right>$
and no even-number product of parity-odd term is constructed.


We consider the parity-even corrections of this theory at first.
For the simplicity, we consider the $\Lambda$-independent terms
which are the leading
in an expansion of $\Lambda$.

The parity-even corrections are distinguished by
how many parity-odd terms are multiplied
when the diagrams are written in the momentum integration form.
The diagrams of (a), (b), (i) and (j) are constructed from
two propagators and two vertices,
0-, 2- and 4-products of parity-odd terms are considerable.
The 0-products do not contain an effect from the CS action,
we only consider the 2- and 4-products here,
but they are excluded from the divergence
by the following reasons.

The 4-products, 
which is constructed by the CS contributions only,
have many $\mu$'s in the numerator.
Such terms do not diverge after the summation of the infinite series
under the region of $\frac{\mu}{M} \ll 1$.
This is a condition that we are working in our calculations%
~\cite{Nittoh:2001full}.

For the products of two parity-odd terms, 2-products,
which come from the crosses of the CS and YM contributions,
the convergent mechanism is different from the above:
all their numerators depend on the external momentum $p$
so the total degree of the internal momentum decreases,
and then divergence does not arise.

%
From the above discussions
the contributions containing an effect of CS action
do not give any divergence,
so it is enough to confirm
the cancellation of the divergence from the pure YM action,
to check the consistency of the regularization method.
Since such divergent contributions are the same as in Ref.~\cite{Nittoh:2000it}
except the space-time dimension,
we can easily confirm that
changing four--dimensional integrations to three--dimensional ones.
%
After a long calculations, we find that
the cancellation mechanism is the same as the quadratic divergence
of four--dimensional YM theory~\cite{Nittoh:2000it},
though the divergence of this case is linear.
Our regularization method also properly works in the YMCS theory.


Next we consider the parity-odd corrections.
%
The procedure of the calculation is almost the same as
Appendix~B in Ref.~\cite{Nittoh:2000it}.
First we treat $\mu$ and $\Lambda$ as finite parameters
but larger than the external momentum $p$, $(\mu,\Lambda\gg p)$.
Then after the Feynman parametrization and the momentum integration,
we see that the parity-odd correction only gives the finite contribution
after the removal of $\mu$ and $\Lambda$
taking the infinite limit of them.

Since we are working with parity-invariant PV pairs,
a parity-odd correction from a PV diagram cancels with its pair
and then all the parity-odd corrections from PV fields
disappear
though all the diagrams considering here give parity-odd corrections.
So the parity-odd corrections only come from the diagrams (a) and (c)
which do not depend on any PV masses.

Following the condition when parity-even corrections are calculated,
taking the limit $\mu, \Lambda \rightarrow \infty$ under 
$\mu/\Lambda\rightarrow 0$,
we get the parity-odd part of vacuum polarization tensor
from (a) and (c) diagrams as
\begin{equation}
\Pi_{\mu\nu}\big|_\mathrm{odd}=
g^2c_v\frac{7}{12\pi}\epsilon_{\mu\rho\nu}p^\rho.
\label{eq:pi-odd mu << Lambda}
\end{equation}
This value corresponds to the result of Refs.%
~\cite{Giavarini:1992xz,Giavarini:1993xb}
where the integer-shift is given
after the renormalization procedure.

%
We worked in the condition $\mu \ll M \ll \Lambda$ in the above calculation.
This condition is necessary for the summation of the infinite series
in the calculation of the divergence
which arises from the parity-even corrections.
For the calculation of the parity-odd corrections,
this condition is not always necessary
because the mass parameter $M$ does not contribute to.
So we can derive the CS shift under the condition $\mu \gg \Lambda$.%
~\footnote{%
  The leading terms are just the same as the ones of the non-YM case
  in the expansion of $\mu$.
  So the divergence can be removed similarly.
  }
%
In such a case, as we see in the below,
the result of infinite limit
under the condition $\mu /\Lambda \rightarrow \infty$
corresponds to the theory without the YM term.

\label{sec:CS+p-even HCD}

We considered YMCS theory so far.
It is not strictly necessary to introduce a YM term as a regulator, however,
if a proper HCD action is introduced and
all the diagrams higher than two-loops are regularized.
As we mentioned in Ref.~\cite{Nittoh:1998ey},
since a general form of a HCD Lagrangian is $FD^nF$
we can consider the theory constructed by the action
without the YM term:
\begin{equation}
S_\Lambda = S_\mathrm{CS}  + S_\mathrm{HCD} + S_\mathrm{GF}^H,
\end{equation}
where all the components are given by
\eqref{eq:CS action}, \eqref{eq:HCD action} and \eqref{eq:GF action}.
Here we calculate the one-loop corrections
in the same manner as the YMCS theory.
%

The regularized action also contains the parameter $\mu$.
It always appear in the form of $\mu\Lambda^2$ in the Feynman rules,
the dependence of the ratio $\mu$ and $\Lambda$
finally disappears in the infinite limit of $\mu$ and $\Lambda$.
So we work in the condition $\mu = \Lambda$ following calculations.


We also work in the region $\Lambda \gg 1$
and only calculate the leading term
in expansion of $\Lambda$
which is independent of $\Lambda$.
%
%
There is no four-point vertex under this condition,
no correction comes from (c) and (d) diagrams.
%
Though the absence of such diagrams,
the cancellation mechanism works similarly in the case of the YMCS theory,
and we also see that all the divergences are cancelled in this theory.
%


For finite corrections,
no diagram other than (a) and (c) gives parity-odd correction
similar to the above calculation of the YMCS theory
because of the parity-invariant PV fields.
%
Here we also take the momentum integration
under a finite $\Lambda \gg p$
and take the infinite limit of $\Lambda$ after the integration.
Under this procedure,
we only pick up the terms which give finite corrections
after the reduction with $\Lambda$'s in the numerator.%
~\footnote{%
 We can show that no divergent corrections arise
 by an easy estimation of the momentum integrations~\cite{Nittoh:2001full}.
 } 
And then we can easily get the finite correction from these diagrams
as follows:
\begin{equation}
\Pi_{\mu\nu}\big|_\mathrm{odd}=
g^2c_v\frac{5}{12\pi}\epsilon_{\mu\rho\nu}p^\rho.
\end{equation}
Here a renormalization procedure is required to obtain the coupling shift%
~\cite{Giavarini:1992xz,Giavarini:1993xb,Chen:1997nv}.
In the same way as the YMCS theory,
we calculate the ghost self-energy and the gauge-ghost-ghost vertex
then the shift is given by
\begin{equation}
\alpha=\frac{11}{9}c_v,
\end{equation}
though the integer-shift is expected
by the universality in Ref.~\cite{Asorey:1994em}
since a parity-even HCD term is dominant in this theory.
As we briefly mentioned above,
this result corresponds to the case of YMCS theory
with the condition $\mu \gg \Lambda$.

\label{sec:conclusions and discussions}

In this paper, we calculate the CS coupling shift
in the framework of the hybrid regularization method
based on a local HCD regulator.
Using with a \textit{parity-invariant} PV regularization,
a parity-even term is dominant at a large momentum limit
and the integer-shift is expected,
but one of our results gives a non-integer-shift $\frac{11}{9}c_v$.
Our results are read that
the integer- or non-integer-shift is governed by
the presence or absence of the YM term,
and a local HCD action may lead non-integer-shift.
%

%
A YM term is usually introduced into the theory but it is not necessary
in a view of the hybrid regularization,
since the main purpose of the HCD term is a regularization of the theory
its form is not so important
so long as the divergence is removed.
It is reasonable to consider the theory without the YM term
as a regularized theory of the CS gauge theory.

%
Though we work in the region $\mu\ll\Lambda$ in the YMCS theory
we also consider the case $\mu\gg\Lambda$ about the shift
because the calculation of the parity-odd part does not depend on
the PV mass parameter.
In this case,
the result of the limit $\mu / \Lambda = \infty$
exactly corresponds to our theory without YM term
including the renormalization procedure.
This fact might suggests that
the YMCS theory of the limit $\mu / \Lambda = \infty$
is equivalent to that theory
and the CS shift varies depending on the ratio between $\mu$ and $\Lambda$.

%
The dependence of the shift on $\mu / \Lambda$
can be derived from the viewpoint of Ref.~\cite{Asorey:1994em}
where the shift is determined by the value of $\phi (\infty)$.
The function $\phi (p)$ is, roughly speaking,
defined by a fractional expression of parity-even and -odd term
in the regularized action as
$\frac{\textrm{(even terms)}}{\textrm{(odd terms)}}$
and in our case $\phi (p) = \frac{\Lambda^2+p^2}{\mu\Lambda^2}p$
where $p$ is the internal momentum of one-loop diagrams.

To remove the divergence,
we have to sum up all the PV regulators by way of Ref.~\cite{Nittoh:2000it,Nittoh:2001full},
and assume that
infinite limit of the internal momentum behaves like mass parameter $M$,
$(p\sim M)$.
In addition to this assumption, for the case of $\mu \ll \Lambda$,
another condition $\mu \ll M$ is needed to sum up all the PV regulators,
and we get $\phi (\infty) = +\infty$
after the removal of $\mu$ and $\Lambda$.
On the other hand for $\mu \gg \Lambda$ or non-YM case,
the condition $\mu \ll M$ is not necessary
so the $\phi (\infty)$ vary depending on the behavior of $M$.

For more exact discussion,
a detailed analysis of the HCD term is necessary
within the framework of the YMCS theory.
We will discuss the role of HCD term in detail
calculating the double scaling limit
of the parameter $\mu$ and $\Lambda$
in Ref.~\cite{Nittoh:2001full}.

%
\providecommand{\href}[2]{#2}\begingroup\raggedright\endgroup

\end{document}